\documentclass[12pt,preprint]{aastex}

\def\ltsima{$\; \buildrel < \over \sim \;$}
\def\simlt{\lower.5ex\hbox{\ltsima}} 
\def\gtsima{$\; \buildrel > \over \sim \;$}
\def\simgt{\lower.5ex\hbox{\gtsima}} 

\begin{document}

\title{Chandra and HST observations of gamma-ray blazars: 
comparing jet emission at small and large scales}

\normalsize \author{F. Tavecchio}
\affil{INAF - Osservatorio Astronomico
di Brera, via Bianchi 46, 23807 Merate (LC), Italy}

\author{L. Maraschi, A. Wolter} \affil{INAF - Osservatorio Astronomico
di Brera, via Brera 28, 20121 Milano, Italy}

\author{C.~C. Cheung\altaffilmark{1}}
\affil{Kavli Institute for Particle Astrophysics and Cosmology, Stanford
University, Stanford, CA 94305, USA}
\altaffiltext{1}{Jansky Postdoctoral Fellow of the National Radio
Astronomy
Observatory.}

\author{R.M. Sambruna} 
\affil{NASA Goddard Space Flight Center, Code 661, Greenbelt, MD 20771, USA}

\author{C.M. Urry} \affil{Yale Center for Astronomy and Astrophysics,
Yale University, 260 Whitney Avenue, New Haven, CT 06520, USA}

\begin{abstract}
We present new {\it Chandra} and {\it HST} data for four
gamma-ray blazars selected on the basis of radio morphology with the
aim of revealing X-ray and optical emission from their jets at large
scales. All the sources have been detected. Spectral Energy
Distributions of the large scale jets are obtained as well as new
X-ray spectra for the blazar cores. Modeling for each object the core
(sub-pc scale) and large-scale (\gtsima 100 kpc) jet SEDs, we derive
the properties of the same jet at the two scales.  The comparison of
speeds and powers at different scales supports a simple scenario for
the dynamics and propagation of high power relativistic jets.
\end{abstract} 

\keywords{Galaxies: active --- galaxies: jets --- (galaxies:) quasars:
individual (0208-512, 0954+556, 1229-021, 2251+158) --- X-rays: galaxies}

\section{Introduction}

In a unified scenario, relativistic jets originating from the accreting
black hole and propagating outwards to kiloparsec scales and beyond must
be present in all radio-loud active galactic nuclei (AGN).  How these jets
form and evolve is not yet known; even intrinsic jet power and composition
are only poorly known. Thus jets remain one of the key puzzles of AGN
astrophysics (see e.g. Blandford 2001, De Young 2002). 

The discovery of X-ray emission from kiloparsec-scale jets -- a major
success of {\it Chandra} (see Harris \& Krawczynski 2006 for a review)
-- has provided us with a new tool to probe jet physics.  Dozens of
new X-ray jets have now been found (see the updated list at {\tt
http://hea-www.harvard.edu/XJET/} and references therein), and
associated {\it Hubble} Space Telescope ({\it HST}) observations have
led to a near doubling of the number of known optical jets (a list can
be found at {\tt http://home.fnal.gov/\verb ~ jester/optjets/}).

The origin of this large-scale emission is probably complex.  Extended
 X-ray emission from low power (FRI) jets is likely synchrotron
 radiation from very high energy electrons connecting smoothly in a
 single emission component to the electrons responsible for the radio
 emission (e.g., Worrall et al. 2001, Kataoka \& Stawarz 2005). On the
 other hand, the spectral energy distributions (SEDs) of the
 multifrequency emission from high power jets in quasars generally
 require two spectral components, as was the case for the first
 discovered extended X-ray jet (PKS 0637-752; Schwartz et
 al. 2000). For instance, in the X-ray survey of selected radio jets
 by Sambruna et al. (2004) which discovered 34 emission regions in
 11/17 jets, only 3 are consistent with a single (power law, or
 convex) synchrotron component while for the other emission regions
 the SED has a concave shape (see also Schwartz et al 2006).  The
 ``rising'' X-ray spectral component has been successfully interpreted
 in a number of cases as inverse Compton (IC) scattering of boosted
 CMB photons by low energy electrons in the jet, implying highly
 relativistic bulk motion up to very large scales (Tavecchio et al.
 2000, Celotti et al. 2001, Siemiginowska et al. 2002, Sambruna et
 al. 2004, 2006a, Cheung 2004, Tavecchio et al. 2004, Schwartz et
 al. 2006).  At this point, other models cannot be discarded, and it
 is also the case that some problems in applying the IC/CMB model to
 specific sources have been pointed out (see Stawarz et al. 2004,
 Atoyan \& Dermer 2004, Kataoka \& Stawarz 2005, Uchiyama et al. 2006,
 Jester et al. 2006, Harris \& Krawczynski 2006 for criticisms and
 alternatives). At the same time, the one-zone IC/CMB model, if
 correct, has the advantage of involving few physical parameters which
 can be constrained by the radio, optical and X-ray emission of
 kiloparsec-scale jets, yielding interesting consequences on the
 physical state of the plasma in the jet and on their dynamics at
 large scales (e.g. Sambruna et al. 2004, 2006a, Marshall et al. 2005,
 Schwartz et al. 2006).

 At the opposite extreme of spatial scales, the innermost region of
 jets is fruitfully investigated through the study of blazars, whose
 emission is interpreted as the relativistically amplified non-thermal
 continuum produced close to the central engine ($d\sim 0.1$~pc) by a
 jet closely aligned to the line of sight (Urry \& Padovani
 1995). Spectral modeling of the full infrared through $\gamma$-ray
 SED (with synchrotron + IC components) yields the electron density
 and energy distribution, magnetic field intensity and bulk Lorentz
 factor of the flow (with some confidence, as shown by the agreement
 between results of different groups; e.g., Ghisellini et al. 1998,
 Tavecchio et al. 2002, Kubo et al. 1998, Sikora \& Madejski 2001),
 allowing us to infer basic global quantities characterizing the jets
 quite close to their origin, in particular the kinetic power and the
 matter content (e.g., Maraschi \& Tavecchio 2003).

Coupling information derived at subparsec- and kiloparsec-scale for
the same jet could have great potential to construct a global
understanding of powerful extragalactic jets.  This approach was first
applied to two well known blazars (1510-089 and 1641+399)
serendipitously belonging to the sample surveyed with {\it Chandra} by
Sambruna et al. (2004). The sample had been selected on the basis of
radio morphology and brightness in order to search for extended X-ray
emission. For both blazars X-ray emission outside the nucleus was
detected.
The results (Tavecchio et al. 2004, hereafter Paper I) were consistent
with a constant bulk Lorentz factor and constant power along the jet
and a simple scaling of the electron density and magnetic field
intensity, suggesting free expansion of the jet between the subpc
(blazar) scale and the ($\sim $100 kpc) scale of the resolved X-ray
knots. Therefore, these results imply that powerful jets are only
weakly affected by the environment up to these scales. Note, however,
that these findings do not exclude that the {\it terminal} portions of
the jets can interact more strongly with the environment, dissipating
part of their power and decelerating. Indeed, there is evidence
(e.g. Georganopoulos \& Kazanas 2004; Sambruna et al. 2006a) suggesting
that, in some cases, the terminal regions of the jet can be affected
by deceleration (possibly due to the cumulative effects of entrainment
of external gas; Tavecchio et al. 2006), accompanied by an increase of
magnetic field intensity and particle density.

 
Unfortunately, only very few jets can be studied on both
scales. Indeed the best-studied inner jets do not tend to have well
studied large-scale jets, precisely because the former are the most
closely aligned with the line of sight, which means projected jet
lengths are small and the large scale jets must be seen in contrast to
the bright, beamed cores. This is why the number of blazars well
observed on pc and kpc scales is very small. It is also worth
noting that in the case of sources displaying large scale jets but
with a weak blazar core, the determination of the parameters
characterizing the small scale jet is less robust, due to the presence
of a mix of jet and disk emission in the core, especially in the crucial X-ray
band (e.g. Sambruna et al. 2006b).  With the aim of increasing the
number of blazars with multifrequency observations of the jet at
large-scales, we proposed {\it Chandra} and {\it HST} observations of
four $\gamma-$ray blazars (0208-512, 0954+556, 1229-021 and 2251+158;
see Tab.1), showing a radio jet suitable for X-ray imaging. The
 detection of the blazar in $\gamma$-rays indicates the presence of a 
strong jet component in the core, generally leading to a more
 reliable estimate of the small scale jet parameters 
 due to the extended sampling of the SED.

In this paper we report the analysis of {\it Chandra} and {\it
HST}\footnote{Based on observations with the NASA/ESA Hubble Space
Telescope obtained at the Space Telescope Science Institute, which is
operated by the Association of Universities for Research in Astronomy,
Incorporated, under NASA contract NAS5-26555.}  data (Sect. 2), the
modeling of the SEDs of the blazar cores and knots in the large scale
jet (Sect. 3), the determination of speeds and powers (Sect. 4) and
the discussion of the results (Sect. 5). Throughout this work we use
the following cosmological parameters: $H_{\rm 0}\rm =70\; km\;
s^{-1}\; Mpc^{-1} $, $\Omega _{\Lambda}=0.7$, $\Omega_{\rm M} = 0.3$.

\section{Observations and data analysis}

Basic characteristics of the blazars analyzed in this work are
reported in Table 1. In the following we report the details of the
procedure used to analyze the data. For the quasar 0208-512 we did not
obtain the requested {\it Chandra} pointing, since it was assigned to
the jet survey of Marshall et al (2005). X-ray and radio data for the
core for this source are directly taken from Marshall et al. (2005)
and Schwartz et al. (2006), while we report only the analysis of our
{\it HST} pointings.

\subsection{Chandra}

The sources were observed with {\it Chandra} (Weisskopf et al. 2000)
with ACIS-S in imaging mode. The journal of observations is reported
in Tab.2. The data were collected with the back-illuminated ACIS-S S3
chip in 1/8 sub-array mode to avoid/minimize the pileup of the central
AGN source, with a frame time of 0.4 s. Telemetry was in faint mode.

The data were reduced with the standard {\it Chandra} pipeline with
the CIAO software (version 3.3) and the most recent available
calibration products (CALDB 2.26), as described in
\verb+http://cxc.harvard.edu/+.  The corrections applied are those
appropriate for the ACIS-S instrument.  No high background periods due
to particle-induced flares were present in the datasets.

Events were selected in the 0.3-10 keV interval for both imaging and
spectral analyses.  The radio and X-ray images were registered
assuming that the cores are coincident. Applied shifts are
0.42$^{\prime\prime}$, 0.22$^{\prime\prime}$, 0.32$^{\prime\prime}$
for 0954+556, 1229-021 and 2251+158, respectively, below {\it Chandra}
aspect uncertainties. The core spectra were extracted in a
1.5$^{\prime\prime}$ region centered on the centroid of the
emission. The jet spectra were extracted in circles centered on bright
radio features (see below). The background was evaluated in nearby
regions devoid of sources. Response matrices are created in the usual
fashion; spectra are binned so that each resulting bin has a S/N $>$
3.

\subsubsection{Cores}

Due to the short frame time the pile-up is negligible in both 0954+556
and 1229-021. Fitting the spectra with a power-law model and free
$N_{\rm H}$ gives a value consistent with the Galactic column density
(from Dickey \& Lockman 1990).  We therefore fit the spectra with
fixed galactic $N_{\rm H}$ and a simple power law.  Results are given
in Table 2. 2251+158 is at least a factor of 10 brighter than the
other two targets, and the effects of pile-up cannot be ignored
(pile-up fraction=29\%).  We use the \verb+pileup+ model in XSPEC
(Davis 2001) to account for the distortion in the spectrum. In this
case we find an $N_{\rm H}$ slightly in excess of the Galactic value.

These sources have been observed several times in the past in X-rays
(Fossati et al. 1998, Wilkes et al. 1994, Siebert et al. 1998, Prieto
1996, Tavecchio et al. 2002, Marshall et al. 2005) with fluxes and
spectral parameters consistent with our results.

\subsubsection{Jets}

Jet knots are quite fainter than the cores and the poor statistics do
not allow to measure the X-ray spectral slope.  Fluxes listed in Table
3 are derived assuming a power-law spectrum of slope $\Gamma=1.7$
(typical for jet knots, e.g. Sambruna et al. 2006a) and Galactic
$N_{\rm H}$.

\subsection{HST}

We observed three of the four blazars with the Advanced Camera for
Surveys (ACS) aboard the {\it Hubble} Space Telescope in two bands. The
remaining source, PKS~1229--021, has existing multi-filter WFPC2 data
available from the archive from which Le Brun et al. (1997) already
detected optical emission in the jet. For our ACS observations, we
observed each target with the F475W and F814W filters (SDSS g and
Broad I, respectively, with effective frequencies of $6.32\times
10^{14}$ Hz and $3.72\times 10^{14}$ Hz) for one orbit per filter. The
Le Brun et al. (1997) data were taken in the F450W and F702W filters,
close to standard B and R bands, respectively, with effective
frequencies of $6.58\times 10^{14}$ Hz and $4.33\times 10^{14}$ Hz.

In 2251+158, Cheung et al. (2005) discovered optical emission from the
jet and hot spot from an archival WFPC2 image. The new ACS images
(characterized by a better resolution) confirm the detections of knots
B and C in the jet (nomenclature from Cheung et al. 2005) and resolve
the hot spot into 2 pieces -- this perhaps means that there is a
previously unresolved portion of the jet in the ``hot spot'' as
reported in Cheung et al. (2005).  Knots A and B are not easily
distinguishable from the wings of the badly saturated optical nucleus
and we could not obtain any useful photometry of them from the ACS
images. Our measurements of the optical jet fluxes in PKS~1229--021
are consistent with those published in Le Brun et al. (1997).

Upper limits of 0.03 $\mu$Jy (3 $\sigma$, in both the F814W and F475W
images) were measured for radio jet knots undetected in the ACS images
(0208--512 and 0954+556). In the archival WFPC2 images of PKS~1229--021,
the corresponding point sources limits are 0.2 and 0.3 $\mu$Jy (3
$\sigma$) in the F702W and F450W images, respectively.

\subsection{Radio}

For 3 of the 4 targets, we obtained and analyzed radio data from the
NRAO\footnote{The National Radio Astronomy Observatory is operated by
Associated Universities, Inc. under a cooperative agreement with the
National Science Foundation.} Very Large Array (VLA) archive using
standard procedures in AIPS (Bridle \& Greisen 1994) and DIFMAP
(Shepherd, Pearson \& Taylor, 1994). We use published ATCA
measurements of PKS~0208--512 from Schwartz et al. (2006).

A single 4.9 GHz image of 0954+556 was analyzed (3.5 min. from program
AH170) to set the normalization for its extended radio emission; Reid
et al. (1995) measured a radio spectral index for the western jet/lobe
complex of $\alpha$=0.9 using data from 0.4--5 GHz and we adopt this
value. For PKS~1229--021, matched resolution 4.8 and 15 GHz data
originally published by Kronberg et al. (1992) were reanalyzed.  We
use published measurements and data for the jet of 2251+158 from
Cheung et al. (2005). For the latter two targets, we created spectral
index maps which show values of $\alpha\sim$0.7 (PKS~1229-021) and
$\alpha\sim$0.8 (2251+158) for the jet.

\subsection{Results: imaging and photometry}

In Figs. 1 (a-c) we show the smoothed X-ray images together with radio
contour overlays. The {\it Chandra} images (plotted in logarithmic
scale) have been smoothed with the \verb+ftools+ tool \verb+fadapt+
using a threshold of 10 counts for 0954+556 and 1229-021 and 20 counts
for 2251+158. Radio contours are plotted logarithmically in steps of
a factor 1.5.

X-ray emission associated to bright radio knots is clearly detected in
all the three sources.  In 0954+556, due to the limited length of the
jet, only the emission associated to the terminal region can be
clearly evaluated. In the jet of 1229-021, X-ray emission tracking the
radio is clearly visible. The brightness of the X-ray emission
decreases and is only barely detected at the terminus. One knot and
the hotspot are detected in 2251+158. In Figs we report the circular
region used to extract the fluxes. Table 3 summarizes the values of
the fluxes measured in the extraction regions centered on bright X-ray
features (for which we use an alphabetical nomenclature, starting with
the region closest to the core)

\section{Modeling the Spectral Energy Distributions}

In Fig. 2 (a-d) we report the SEDs, for the cores (upper panels) and
emission regions in the large scale jet (lower panels), constructed using data
from the historical records and the {\it Chandra} and {\it HST} data
presented in this work. Historical data used to construct the blazar
SEDs are taken from the references reported in Tavecchio et al. (2002)
for 2251+158 and 0208-512, while those for 0954+556 and 1229-021 are
reported in the figure caption. It is worth noting that the blazar
nature of 0954+556 and its identification with the EGRET source
3EG~J0952+5501 has been recently questioned on the basis of new VLBA
images revealing a Medium Symmetric Object morphology
(Rossetti et al. 2005; see also Marscher et al. 2002). This is contrary 
to the typical morphology of blazars, characterized by the presence of
strong, compact features. It is therefore more likely that the
counterpart to 3EG~J0952+5501 is J0957+5522 (Sowards-Emmerd et
al. 2003), not 0954+556 as originally suggested by Mattox et
al. (2001).

All these blazars are well studied, with several observations in the
X-ray band. For clarity we only report the new {\it Chandra} data. We
stress that all these data are not simultaneous and that the sources
can undergo large variations. In this respect, the best example of a
variable source among those considered here is 2251+158, which in the
spring 2005 displayed a strong outburst, with a large (at least a
factor of 3) increase of the optical and the hard X-ray luminosities with
respect to quiescent levels (Pian et al. 2006)

\subsection{The Blazar region}

We modelled the SEDs of the inner jet with the emission model fully
described in Maraschi \& Tavecchio (2003), considering synchrotron and
IC (both Synchrotron Self-Compton and External Compton) radiation.

To reproduce the observed shape of the two humps in the SED we assume
that the electron energy distribution is described (between $\gamma
_{\rm min}$ and $\gamma _{\rm max}$) by a smoothed broken power law
with indices $n_1$ and $n_2$ below and above the break located at
$\gamma _b$. This purely phenomenological description accounts for the
observed shape of the synchrotron and IC components. In the widely
assumed diffusive shock acceleration model (e.g. Kirk et al. 1998) or
in cases of severe cooling one would expect $n_1=2$. However, there
are several objects for which this limit seems to be violated, with
values as extreme as $n_1=1.4-1.5$ (e.g., Piconcelli \& Guainazzi
2005, Yuan et al. 2005). It is conceivable that, at least in these
cases, the electron distribution derives from a (continuously
operating) different acceleration mechanism (for possibilities, see
e.g, Sikora et al. 2002; Katarzy{\'n}ski et al. 2006).

Relativistic electrons and tangled magnetic field with intensity $B$
fill the source, assumed to be a sphere with radius $R$. Relativistic
effects are described by the relativistic Doppler factor $\delta $,
given by $\delta = [\Gamma (1-\beta \cos \theta)]^{-1}$, where $\beta
=v/c$, $\Gamma $ is the bulk Lorentz factor and $\theta $ is the
viewing angle with respect to the blob velocity.

In the SED we also include a black body with luminosity $L_{\rm disk
}$ and temperature $T=10^4$ K, intended to provide a crude
representation of the blue bump originating in the accretion disk. A
fraction $L_{BLR}=\tau L_{\rm disk}$ of this radiation is thought to
be reprocessed by the Broad Line Region clouds, located at a distance
$R_{BLR}$ from the central BH, providing the source of the external
photons for the EC process.

Estimates of $L_{BLR}$ for 0954+556, 1229-021 and 2251+158 are given
by Cao \& Jiang (1999). For 0208-512 we estimate $L_{BLR}$ using the
methods of Celotti et al. (1997) and the flux of the MgII line
provided by Scarpa \& Falomo (1997). Given $L_{BLR}$, we fix $L_{\rm
disk}$ assuming the parameter $\tau $ fixed to 0.1, a value consistent
with the ratio between $L_{BLR}$ and $L_{\rm disk}$ inferred for the
(few) radio-loud sources for which a measure of both quantities is
simultaneously available (e.g. Sambruna et al. 2006b).

In Fig. 2 (top panels) we report the total emission (solid line) and
the separate contributions from the different spectral components
(synchrotron: dotted; SSC: short dashed; EC: long dashed; disk:
dot-line). In all cases the measured slope of the X-ray continuum
appears to be hard but not as extreme as in other powerful blazars
(where $\alpha _X <0.5$, e.g. Tavecchio et al. 2002), suggesting an
important contribution of the (soft) SSC component which, typically,
peaks close to the X-ray band. 

The parameters used to calculate the model (reported in Tab. 4) are
similar to those usually found for this type of sources (e.g. Maraschi
\& Tavecchio 2003). In all cases we fix the value of the minimum
Lorentz factor of the emitting electrons at $\gamma _{\rm min}=1$, as
derived in the case of other powerful blazars (e.g. Maraschi \&
Tavecchio 2003). However, this was not possible in the case of
0954+556, because of the relatively steep X-ray continuum ($\alpha
_X\sim 1$), which favors a dominant contribution of the SSC emission
and a minor contribution from the (flatter) EC component. In this case
we use $\gamma _{\rm min}=8$. However, we recall that the association
with the EGRET emission is questionable, and therefore the results for
the latter source should be considered with caution. For all the
sources we used a value of $R_{BLR}$ consistent with the the
correlation between $R_{BLR}$ and $L_{BLR}$ found by Kaspi et
al. (2005) and Bentz et al. (2006). 

Although the sampling of the SEDs is good only in the case of
2251+158, the data constrain the model parameters sufficiently for the
present purpose. In fact the lack of data for some of the objects in
the $10^{11}-10^{14}$ Hz region leaves rather unconstrained a portion
of the synchrotron continuum which does not influence the derivation
of the most important parameters. In particular, the power critically
depends on the number of particles carried by the flow (see Sect. 4),
which, due to the steep electron distribution, is constrained by the
well defined X-ray emission, dominated by EC radiation from low-energy
electrons. Note also that the intensity of the $\gamma$-ray emission
determines the radiative output (i.e. the radiative efficency) of the
jet but has a limited influence on the derivation of the kinetic
power, since $\gamma$-ray emitting electrons do not contribute
significantly to the total number of particles.

\subsection{The Large-scale jet emission}

We model the emission using the IC/CMB model, already used in Paper
I. The emitting region is modeled as a sphere with radius $R$, filled
by relativistic electrons and tangled magnetic field with intensity
$B$, in motion with Lorentz factor $\Gamma $. Electrons follow a
power-law law in energy with index $n$, $N(\gamma)=K \gamma ^{-n}$,
between the extremes $\gamma _{\rm min}$ and $\gamma _{\rm max}$ and
emit through synchrotron and Inverse Compton emission. For the latter
it is assumed that the dominant soft radiation field is the CMB. To
uniquely determine the parameters we assume equipartition between the
relativistic electrons and the magnetic field (see Sambruna et
al. 2006b for a discussion).

The radius assumed in the models corresponds to the angular size of
the circular regions used to extract the fluxes. As suggested by the
radio images, with this choice we are probably overestimating the
actual emitting volume. However, it is easy to show that the derived
parameters depends rather weakly on the assumed volume (see e.g. the
Appendix of Tavecchio et al. 2006). For instance, with a the volume
equal to 1/10 of the value assumed here, both the magnetic field and
the Doppler factor would be larger by a factor $\sim 1.3$. Similarly,
the impact of the assumed size on the derive power is rather minor. In
the case of 0208-512, for which Schwartz et al. (2006) use a
rectangular extraction region (and therefore model the emission region
as a cylinder), we derive an effective radius, such that the volume of
the sphere is equal that of the cylindrical region. Apart for the case
of 0208-512 we fix the slope of the electron energy distribution using
the radio spectral index derived above (Sect. 2.3).  The 0208-512
radio jet spectral index of 0.8 is assumed, as the actual measurements
over the short frequency baseline (4.8-8.6 GHz) gave unreliable values
(Schwartz et al. 2006).  The value of the minimum Lorentz factor of
the electrons has been chosen so that the break in the IC continuum is
located between the optical and the soft X-ray band. Of course the
choice of this value is not unique, since values of $\gamma _{\rm
min}$ in the range 5--30 are usually allowed by the data. A deeper
discussion of this point is reported in the next paragraph.

In Fig. 2 (lower panels) we show the synchrotron-IC/CMB spectral
models derived assuming the input parameters reported in Tab.5.

For 0954+556 the X-ray flux could only be extracted for the region
corresponding to the terminal portion of the jet. As discussed in
Tavecchio et al. (2005), also in this case a beamed IC/CMB model
appears necessary, however with
a Doppler factor lower than typically found for knots in the
jet.

For 1229-021 we can model the emission at three different locations
along the jet. Interestingly, we can reproduce the SEDs of the three
regions decreasing only the value of the Doppler factor (from $\delta
\simeq 9$ to $\delta \simeq 5$), and maintaining all the other
quantities (almost) constant. As well, the clear bending of the jet
between regions A and B suggests a change in the (projected) direction
of the jet speed. It is tempting to associate this change with an
increase of the jet angle, which could simultaneously explain the
decrease of the observed beaming (while a ``true'' deceleration would
imply also the increase of both the magnetic field intensity and the
particle density).

\section{Speed and power}

The parameters derived by reproducing the SEDs can be used to infer
the bulk Lorentz factor and the total power characterizing the jet.
In the case of large scale jets with more than one emission region, we
calculate the power and the speed at the region closest to the
core. With this choice we avoid the portion of the jet near the
terminal region which could be affected by deceleration. For the same
reason we do not include 0954+556 in this analysis, since we do not
have multi frequency information on emitting regions before the jet
terminus.

For both the inner, unresolved region and the (still relativistic)
resolved jet, the transported power can be estimated using the
expression:\footnote{As pointed out in Schwartz et al. (2006), this
expression, neglecting the contribution of the pressure, is not
completely correct. However, it provides values differing only by a
small factor from the correct expression and we can therefore safely
use it in this work in which we consider order of magnitude
estimates.}
\begin{equation}
P_{\rm j} = \pi R^2 \Gamma ^2 ( U^{\prime 
    }_B+U^{\prime }_e+U^{\prime }_p)c 
\end{equation}
\noindent
(Celotti \& Fabian 1993), where $R$ is the radius of a cross section
of the jet, $U^{\prime }_e$, $U^{\prime }_p$, and $U^{\prime }_B$ are
the rest frame energy densities of relativistic electrons, protons,
and magnetic field, respectively. 
In the following we assume a plasma composition of one proton per
electron (see e.g. Paper I for a discussion of this choice). With this
choice the total power is largely dominated by the kinetic power
associated to the protons.

As already discussed in Paper I, in order to derive from the Doppler
factor, directly provided by the emission models, the bulk Lorentz
factor of the flow, an observing angle has to be assumed. For the
blazar jet, as in Paper I, we assume that the value of the angle falls
in the range $\theta _{\rm max}$ to 0, where $\theta _{\rm max}=1/\delta$
is the {\it maximum} angle allowed by a given value of the Doppler
factor. The corresponding range covered by the bulk Lorentz factor
will be $\delta-\delta/2$. In the case of the large scale jet,
instead, due to the larger uncertainty in the derived Doppler factor
(see below) for simplicity we decided to assume in all the cases the
maximum angle $\theta _{\rm max}=1/\delta$.

A meaningful comparison between between jet speeds and powers at
different locations in the same jet requires an estimate of the
uncertainties affecting the physical quantities derived by the
spectral fitting.  In the case of the blazar cores, the uncertainties
in the extended multiwavelength coverage of the SEDs generally
constrains the parameters of the model adequately.  We therefore
compute speeds and powers fixing the parameters derived from the
spectral modeling (Tab.4) and consider only the uncertainty in the
Lorentz factor due to the unknown viewing angle, giving for $\Gamma$
the range $\delta/2 < \Gamma <\delta$.  For the large scale jets, the
poor spectral coverage allows relatively large uncertainties in all
the derived quantities. In particular, the data do not strongly
constrain the value of $\gamma _{\rm min}$. However, it is relatively
easy to assess these uncertainties since the IC/CMB model allows a
simple analytical formulation (Schwartz et al. 2006, Tavecchio et
al. 2006).  Using the analytical relations reported in Tavecchio et
al. (2006) we calculate the parameters $\delta $, $K$ and $B$ ,
varying $\alpha$ and $\gamma _{\rm min}$ within the range allowed by
the data and taking also into account the error on the measured
fluxes, dominated by the X-ray one (for 0208-512, for which
uncertainties on the fluxes are not provided in Schwartz et al. 2006,
we assume a typical error of 25\% on the X-ray flux).  For 1229-021
and 2251+158 $\alpha$ is varied from $\alpha _r-0.05$ to $\alpha
_r+0.05$ while for 0208-512 the (conservative) range 0.5--0.8 is
considered. The value of $\gamma _{\rm min}$ is free to vary in the
range 1-100\footnote{Note that not all the combinations of the
parameters are acceptable, since constraints on the position of the
low-energy break of the IC component translates into limits for
$\gamma _{\rm min}$ and $\delta $. In fact, the break of the IC
component, approximately given by $\nu _{\rm break}=\nu _{CMB} \delta
^2 \gamma _{\rm min}^2$, must be located between the optical-UV band
and the soft X-ray band, {\it i.e.} we keep only the combination of
the parameters satisfying $10^{15} \, {\rm Hz} < \nu _{\rm break}<
2.4\times 10^{16} \,{\rm Hz}$.}. We characterize the uncertainty on
$\delta $ and $P_j$ taking the range spanned by the two parameters in
all the possible realizations.

The results are reported in Table 6. We also report the results for
the two blazars already studied in Paper I (1510-089 and 1641+399),
for which we recalculate the uncertainties accordingly to the
procedure discussed above. A comparison between Lorentz factors and
powers determined for the the blazar core ({\it inner}) and the large
jet knots ({\it outer}) is reported in Fig.3, in which the rectangles
include all the values of $\Gamma$ (left) and $P_j$ (right) allowed by
the data. In this plot we also report the limits derived for the two
sources analyzed in Paper I, 1510-089 and 1641+399, in which the
uncertainties have been recalculated using the same procedure above.

The plots show that, on average, the Lorentz factor and the power
derived at the two scales are in agreement, confirming and extending
the results of Paper I. However, the large uncertainty, affecting in
particular the derived power (in particular the values of $P_{\rm
outer }$, spanning in some cases a range larger than a decade),
prevent to draw a stronger conclusion.



\section{Discussion}

The possibility to simultaneously probe the inner and outer part of a
jet offers the unique opportunity to follow its structure over many
orders of magnitude in distance from the central engine. The analysis
presented here confirms and extends the previous findings of Paper I,
obtained for a smaller sample of two sources, indicating the
consistency of speeds and powers independently estimated at the two
scales.
Similar results, showing the consistency of the speeds independently
inferred at VLBI scale and at large scale, have been reported by
Jorstad \& Marscher (2004, 2006). 

Admittedly, the present results are based
on specific assumptions regarding the origin of the large scale jet
emission and the conditions in the jet (i.e. equipartition). In
particular, the estimate of the power critically relies on the
determination of the number of protons in the jet which is
estimated from the low energy end of the relativistic electron
distribution i.e. from the minimum Lorentz factor,$\gamma_{\rm min}$.
In fact $\gamma_{\rm min}$ is well constrained in the framework of 
the IC/CMB model through the discontinuity between the optical and 
X-ray emission. 

If the X-ray emission of the large scale jet is interpreted as 
synchrotron radiation from a second population of relativistic
electrons (e.g. Stawarz et al. 2004, Uchiyama et al. 2006)
$\gamma_{\rm min}$ cannot be constrained since the low-energy 
electrons emit well below the observed radio frequencies.
In this case, lower jet powers can be derived simply through a 
choice of a large value for $\gamma _{\rm min}$.
The lower requirement for the power is sometimes
used as an argument supporting the synchrotron interpretation against
the IC/CMB model, (see e.g. Atoyan \& Dermer 2004): in this case one
should assume that most of the jet power is dissipated in the
propagation of the jet from the small to the large scale

Also within the IC/CMB interpretation, in some cases, there is
complementary evidence suggesting that before its termination the jet
suffers important deceleration, marked by a decreasing Doppler factor
and the increase of magnetic field intensity and particle density
(Georganopoulos \& Kazanas 2004, Sambruna et al. 2006a). The
deceleration can be plausibly induced by entrainment of external gas
(Tavecchio et al. 2006), whose effects become important only when the
cumulative amount of entrained gas reaches some appreciable
level. Moreover, the mixing layer thought to permit the entrainment of
the gas into the jet is believed to grow along the jet (e.g. De Young
2006). Therefore, entrainment can coexist with the evidence of the
conservation of power and speed, since the deceleration is expected to
become important only after some distance along the jet. Therefore all
these elements can be unified in a simple scenario, in which very
powerful jets evolve freely, almost unperturbed, up to large ($\sim
$100 kpc) scale, conserving the original power and speed. In some
cases (depending on external conditions and jet power), the entrained
mass becomes dynamically important before the jet end, leading to the
inferred deceleration (e.g. the case of 1136-135, Tavecchio et
al. 2006).



The detection of resolved X-ray emission from the jets associated with
the $\gamma$-loud blazars studied in this work adds to previous
evidence that resolved X-ray emission from radio-loud quasars with
bright radio jets is common. It is interesting to note that
large-scale X-ray emission has been detected in all the {\it
$\gamma$-ray blazars} with a radio jet long enough to be resolvable
and observed by {\it Chandra}. The present list comprises 11 objects
(10 if we exclude the possible misidentified 0954+556): besides the 2
blazars discussed earlier and the 4 studied here, it includes 3C273
(Sambruna et al. 2001, Marshall et al. 2001), 3C279 (Marshall et
al. 2003), 0827+243 (Jorstad \& Marscher 2004), 1127-145
(Siemiginowska et al. 2002) and 1222+216 (Jorstad \& Marscher
2006). All show conspicuous X-ray emission, with a detection rate of
100\%, even larger than that for sample containing other kinds of
quasars (Sambruna et al. 2004, Marshall et al. 2005).  This fact is
consistent with the idea that beaming plays an important role in
determining the observed X-ray emission (i.e. large scale jets
associated to blazars are likely aligned with the observer), although
the small numbers involved prevent a firm statistical conclusion. {\it
GLAST} (scheduled to be operative at the end of 2007) is expected to
greatly enlarge the number of known $\gamma$-ray radio-loud
AGNs. Likely, most of the sources associated to large scale jets with
known X-ray emission will be detected in the $\gamma$-ray band,
allowing to better characterize the SED of the core and therefore
increasing the number of object suitable for the study.

\acknowledgments FT, AW and LM acknowledge support from grant ASI-INAF
I/023/05/0. Support for Proposal number HST-GO-10004.01-A was provided
by NASA through a grant from the Space Telescope Science Institute,
which is operated by the Association of Universities for Research in
Astronomy, Incorporated, under NASA contract NAS5-26555.
This research has made use of the NASA/IPAC Extragalactic Database
(NED) which is operated by the Jet Propulsion Laboratory, California
Institute of Technology, under contract with the National Aeronautics
and Space Administration.

\newpage

\vskip 1.5 truecm 
 
\centerline{ \bf Figure Captions} 
 
\vskip 1 truecm 
 
\figcaption[images]{X-ray and radio images of 0954+556, 1229-021 and
2251+158. Clearly, X-rays and radio knots in the jets roughly
coincide.}
 
\figcaption[seds]{Spectral Energy Distributions for cores (upper
panel, non-simultaneous data) and jet features (lower panel) of the
sources discussed in the text. Historical data are taken from: K{\" u}hr et
al. (1981), Bloom et al. (1994), Gear et al. (1994), Pian, Falomo \&
Treves (2005) and NED for 0954+556; K{\" u}hr et al. (1981), Tornikoski et
al. (1996), Stickel et al. (1994), Pica et al. (1988), Pian, Falomo \&
Treves (2005) and NED for 1229-021. References for the SED of 2251+158
and 0208-512 are reported in Tavecchio et al. (2002). SEDs of jet
knots show the typical two component structure, well explained by the
IC/CMB model. Note that HST data, although mostly non-detections,
provide very important limits that prevent a single-component
interpretation of the SED.}
 
\figcaption[power]{Comparison between the jet Lorentz factor (left)
and the jet power (right) independently evaluated for the core and the
kpc-scale jet. Together with the sources discussed in the paper we
also report the results for 1510-089 and 1641+399 adapted from Paper
I. The jet speed appears roughly constant, while the jet power could be
constant, although the large uncertainties prevent a firm conclusion.}

\newpage 


\begin{table}[h]
\begin{center}
\begin{tabular}{lcccc}\hline\hline
Source &$z$ & $F_{5 GHz}^{\rm core}$ & $F_{5 GHz}^{\rm jet}$ & $N_{H, gal}$\\ 
&    &   Jy   & mJy   & $10^{20}$ cm$^{-2}$  \\\hline
0208-512 (PKS 0208-512) & 1.003 & 2.1 & 384 & 2.94 \\
0954+556 (4C+55.17)& 0.901 & 1.6 & 91 &  0.90   \\
1229-021 (4C-02.55) & 1.045 & 0.6 & 218 & 2.28 \\
2251+158 (3C454.3)& 0.86& 10.9& 276 & 6.50 \\\hline
\end{tabular}
\end{center}
\caption{\scriptsize The Targets. Notes: 1=Source IAU name;
2=Redshift; 3=Core flux at 5 GHz (in Jy); 4= Total jet Flux at 5 GHz
(mJy); 5= Galactic hydrogen column density (Dickey \& Lockman 1990).}
\label{sources}
\end{table}

\begin{table}[h]
\begin{center}
\begin{tabular}{lccccccc}
\\
\hline
\hline
Source & Date  & Seq. & Exp. time& \, \, $\Gamma $\, \, & $N_H$ & $F_{[2-10\, \rm keV]}$ & $\chi^2/$d.o.f.\\ 
      & & & ksec & & 10$^{20}$ cm$^{-2}$ & $10^{-12}$ erg cm$^{-2}$ s$^{-1}$  &\\
\hline
\hline
0954+556& 6/16/2004 & 700924  &  34.4 & 1.92$^{1.96}_{1.90}$ & fix & 0.38$\pm$0.02 & 1.17/301\\
1229-021& 4/18/2004 & 700923 &  18.3 & 1.61$^{1.66}_{1.57}$ & fix & 1.05$\pm$0.03 & 1.09/151 \\
2251+158 &8/4/2004 & 700925 &  16.4 & 1.63$^{1.66}_{1.58}$  & 9.4$^{10}_{8.2}$  & 12.9$^{14.2}_{11.9}$  & 0.97/412 \\ 
\hline
\end{tabular}
\end{center}
\caption{\scriptsize Journal of {\it Chandra} observations and
parameters of the best fit with absorbed power-law model of the the
cores. Quoted errors are at the 90\% confidence level for 1 parameter
of interest.}
\label{cores}
\end{table}

\begin{table}[h]
\begin{center}
\begin{tabular}{lcccccc}\hline\hline
Source & knot & $F_{\rm 5 GHz}$ & $F_{\rm opt,1}$ & $F_{\rm opt,2}$ & $F_{\rm 1 keV}$ \\ 
&    &   mJy   & $\mu$Jy   &  $\mu$Jy   & nJy \\\hline
0208-512 & R1  & 12.8$^a$ & $<0.03$ & $<0.03$ & 3.3\\
         & R2  & 25.5$^a$ & $<0.03$ & $<0.03$ & 4.7\\
0954+556 & A (HS) & 79$\pm$8 & $<0.03$ & $<0.03$ & 0.3$\pm0.1$\\
1229-021 & A & 107$\pm$11 &  $0.47\pm0.21$ & $<0.3$& 8.5$\pm3.2$\\
         & B& 19 $\pm$2 & $<0.2$ & $<0.3$& 1.1$\pm0.4$\\
         & C (HS?)& 39$\pm$4 & $<0.2$& $<0.3$& 1.3$\pm0.4$\\    
2251+158 & A & 44$\pm$4 &0.23$\pm$0.04 & 0.08$\pm$0.03 & 6.0$\pm$1.4\\ 
         & B (HS)   & 216$\pm$22 &0.62$\pm$0.04 & 0.43$\pm$0.06 & 6.0$\pm$1.4\\ \hline
\end{tabular}
\end{center}
\caption{\scriptsize Multifrequency fluxes of the jet knots. Notes:
$^a$ Fluxes measured at 8.6 GHz, taken from Schwartz et
al. (2006). $F_{opt,2}$ measured at effective frequencies of
$3.72\times 10^{14}$ Hz and $6.32\times 10^{14}$ Hz (filters F814W and
F475W) for 0208-512, 0954+556 and 2251+158 and $4.33\times 10^{14}$ Hz
and $6.58\times 10^{14}$ Hz (filters F702W and F450W) for 1229-021.}
\label{knots}
\end{table}

\begin{table}[h]
\begin{center}
\begin{tabular}{lcccccccccc}
\\ 
\hline 
\hline 
Source & $R$ & $B$ & $\delta$ & $\gamma _{\rm b}$ & $\gamma _{\rm max}$&$n_1$& $n_2$ & $K$ & $L_{\rm BLR}$ & $R_{\rm BLR}$\\
& $10^{16}$ cm& G& & & $10^4$ & & & cm$^{-3}$ & $10^{45}$ erg s$^{-1}$ &$10^{17}$ cm \\ 
\hline
0208-512 & 1.5& 2.5 & 18& 150& 1&1.6& 3.8& $1.2\times 10^4$& 2.7 & 4.8\\
0954+556 & 1.5& 2.5 & 12& 100& 1.5& 2& 3.5& $1.2\times 10^5$& 0.9& 3.3\\
1229-021 & 2& 3.2 & 10& 200& 1&1.4& 3.5& $5.8\times 10^4$& 4.5& 7.5\\
2251+158 & 4& 1.8 & 12& 115& 2&2& 3.4& $5\times 10^4$& 4& 9.5\\\hline
\end{tabular}
\end{center}
\caption{\scriptsize Parameters used for the blazar emission model
described in the text. The minimum Lorentz factor of the electrons,
$\gamma _{\rm min}$, is fixed to 1 for all the sources, but the case of
0954+556 for which $\gamma _{\rm min}=8$.}
\label{blazarfit}
\end{table}   
\begin{table}
\begin{center}
\begin{tabular}{lcccccccc}
\\ 
\hline 
\hline 
Source & Knot & n & B & K & $\delta $& $\gamma _{\rm min}$ & $\gamma _{\rm max}$ & $R$\\
 & & &$\mu$G &$10^{-6}$ cm$^{-3}$ & & & $10^{5}$ & $10^{22}$ cm \\ 
\hline
0208-512 & R1$^a$ & 2.6 & 8 & 8 & 11.8 & 10& 3& 1.8 \\
         & R2$^a$ & 2.6 & 9&  15& 11 &   10 &3 &1.8  \\
0954+556 & A & 2.8 & 38& 450& 3.8& 20& 1& 1.9 \\
1229-021 & A& 2.4 & 10 & 10& 8.6& 25& 4& 2.2 \\
         & B& 2.4 & 10 & 8& 6& 25& 4& 2.2 \\
         & C & 2.4& 10& 8.5 & 5.4& 20& 4& 2.9 \\
2251+158 & A& 2.5& 7& 5& 10& 14& 5& 2.3 \\
         & B & 2.6& 14.5& 20& 8& 7& 3& 2.83 \\ 
\hline
\end{tabular}
\end{center}
\caption{\scriptsize Parameters used for the IC/CMB model for the
large scale jets reported in Fig.2. $^a$: we maintain the original
nomenclature for the regions in the jet adopted by Schwartz et
al. (2006).}
\label{blazarfit}
\end{table}


\clearpage

\begin{table}
\begin{center}
\begin{tabular}{lcccc}
\\ 
\hline 
\hline 
Source &  & $\delta$  & $\Gamma$ & $P_{\rm jet}$ \\
&&&& $10^{47}$ erg/s \\
\hline
0208-512  & {\it inner} & 18 & 18--9& 2.2--0.55 \\
         & {\it outer} & 11.8--5.5 & 11.8--5.5& 7.9--0.16 \\
1229-021  & {\it inner} & 10 & 10--5& 4.8--1.2 \\
          & {\it outer} & 7.9--4.7& 7.9--4.7 & 16--1 \\
1510-089$^a$  & {\it inner} & 19 & 19--9.5& 5--1.25 \\
          & {\it outer} & 16.7--10& 16.7--10  & 2.8--0.1 \\
1641+399$^a$  & {\it inner} & 9.7 & 9.7--5& 1.2--0.3 \\
          & {\it outer} & 7.1--4.5& 7.1--4.5  & 0.5--0.1 \\
2251+158  & {\it inner} & 12 & 12--6& 25--6.2 \\
          & {\it outer} & 7.7--5.3 & 7.7--5.3& 12--0.75 \\ \hline 
\end{tabular}
\end{center}
\caption{\scriptsize Derived parameters for inner and outer
jets. 0954+556 is excluded, since only information on the terminal
region is available. $^a$Data for 1501-089 and 1641+399 are adapted from
Paper I.}
\label{power}
\end{table}    

\clearpage


\begin{figure}[]
\figurenum{1a}
\noindent{\plotone{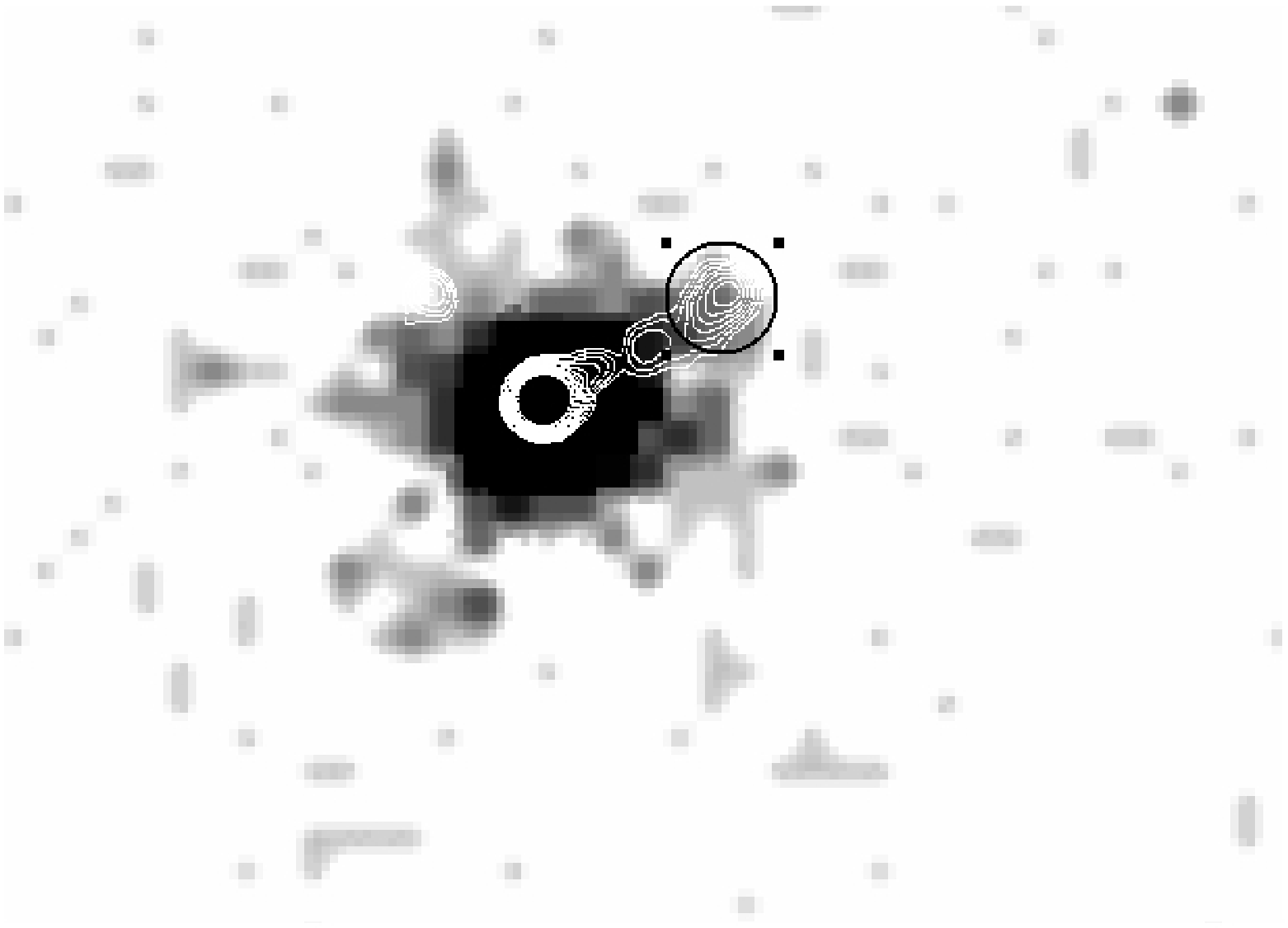}}
\caption{}
\end{figure}

\begin{figure}[]
\figurenum{1b}
\noindent{\plotone{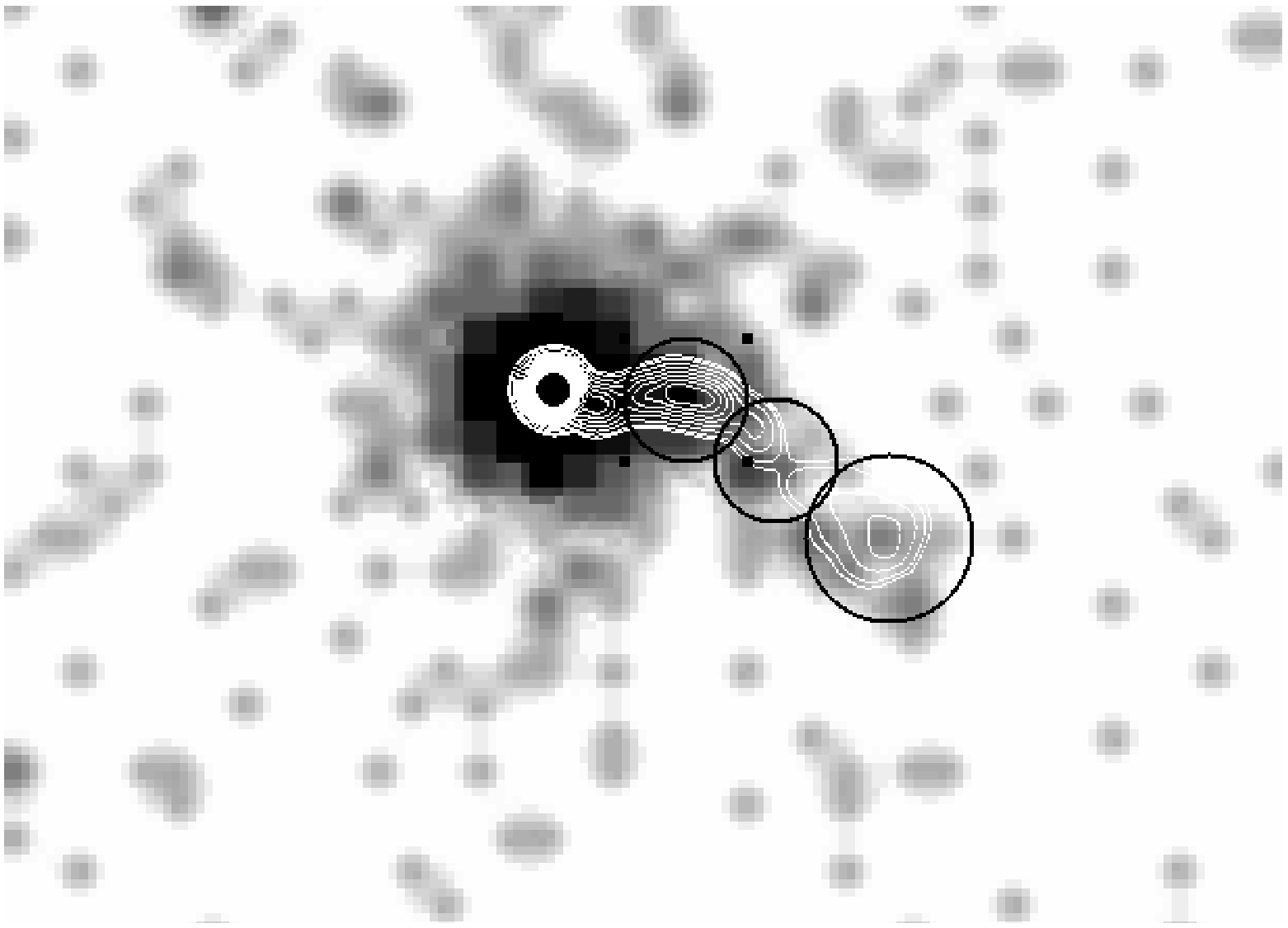}}
\caption{}
\end{figure}

\begin{figure}[]
\figurenum{1c}
\noindent{\plotone{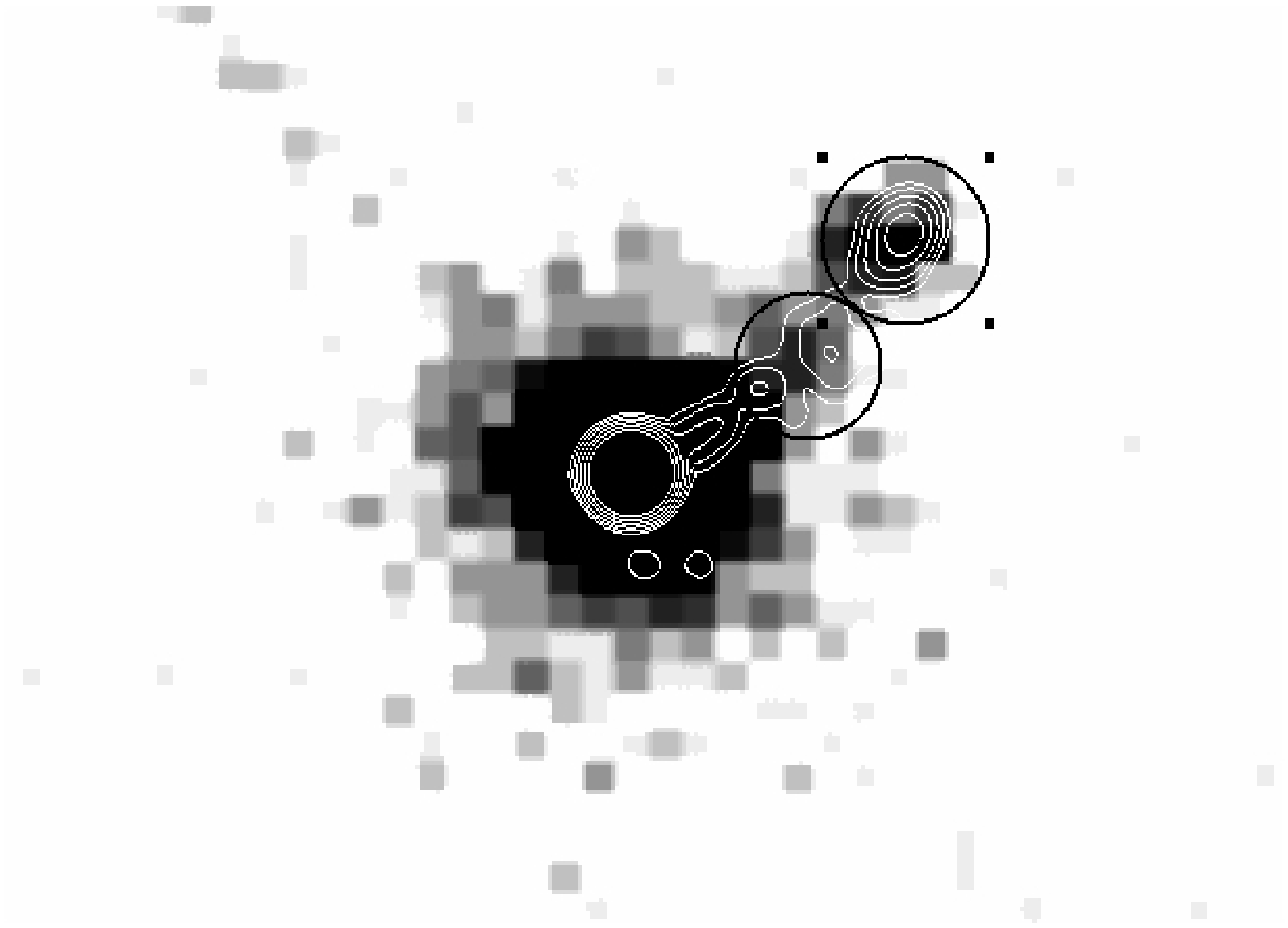}}
\caption{}
\end{figure}

\begin{figure}[]
\figurenum{2a}
\noindent{\plotone{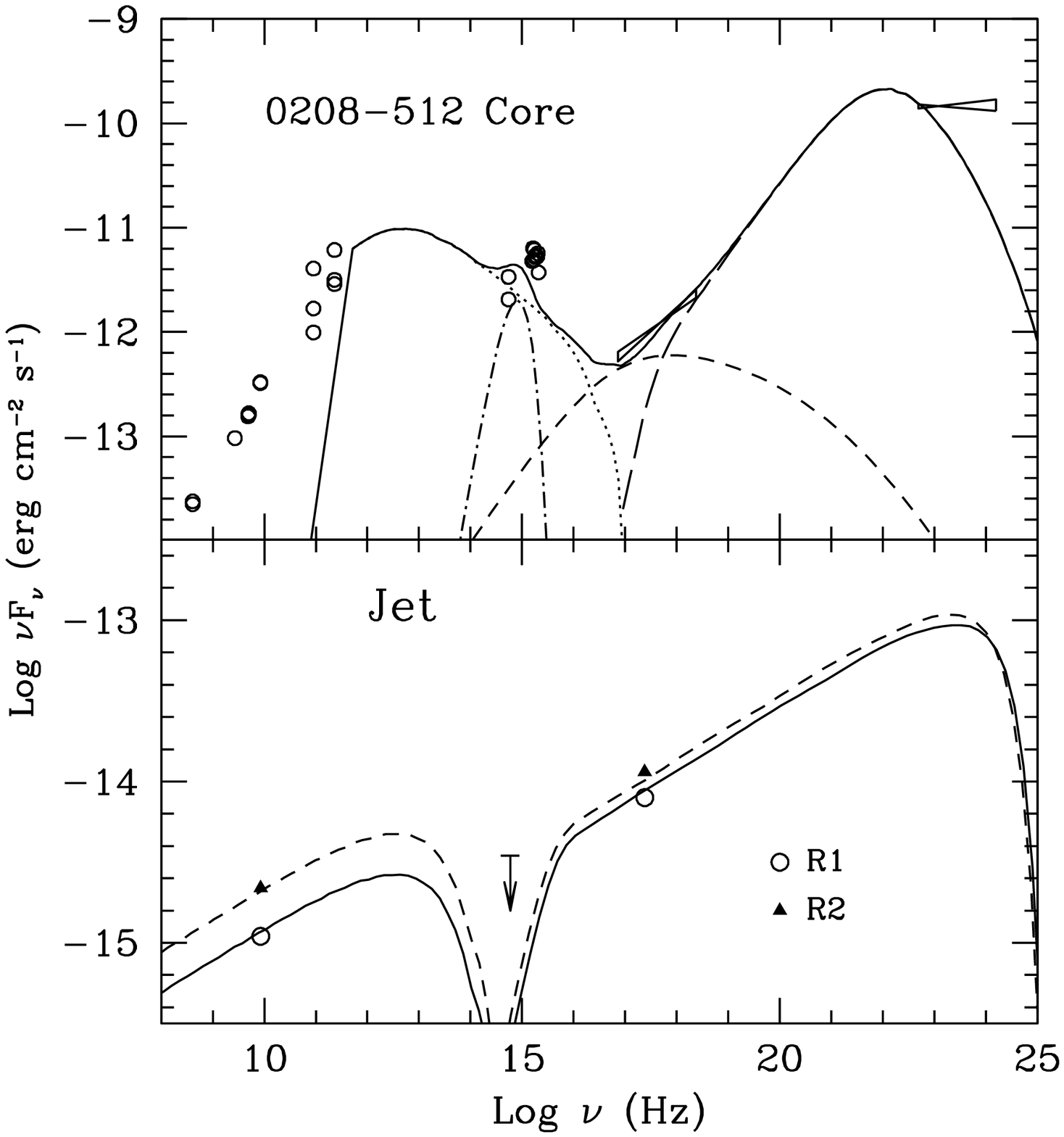}}
\caption{}
\end{figure}

\begin{figure}[]
\figurenum{2b}
\noindent{\plotone{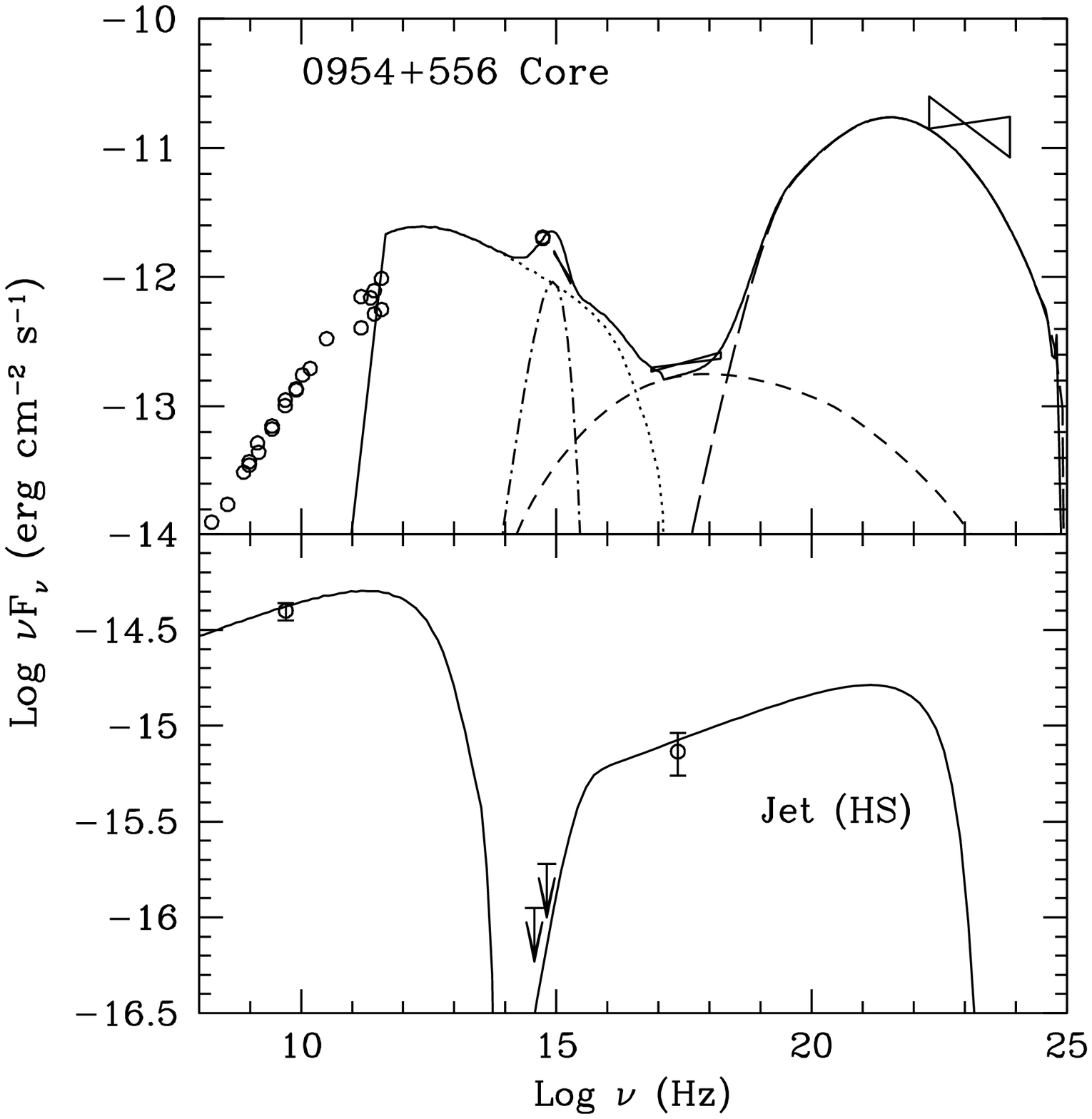}}
\caption{}
\end{figure}

\begin{figure}[]
\figurenum{2c}
\noindent{\plotone{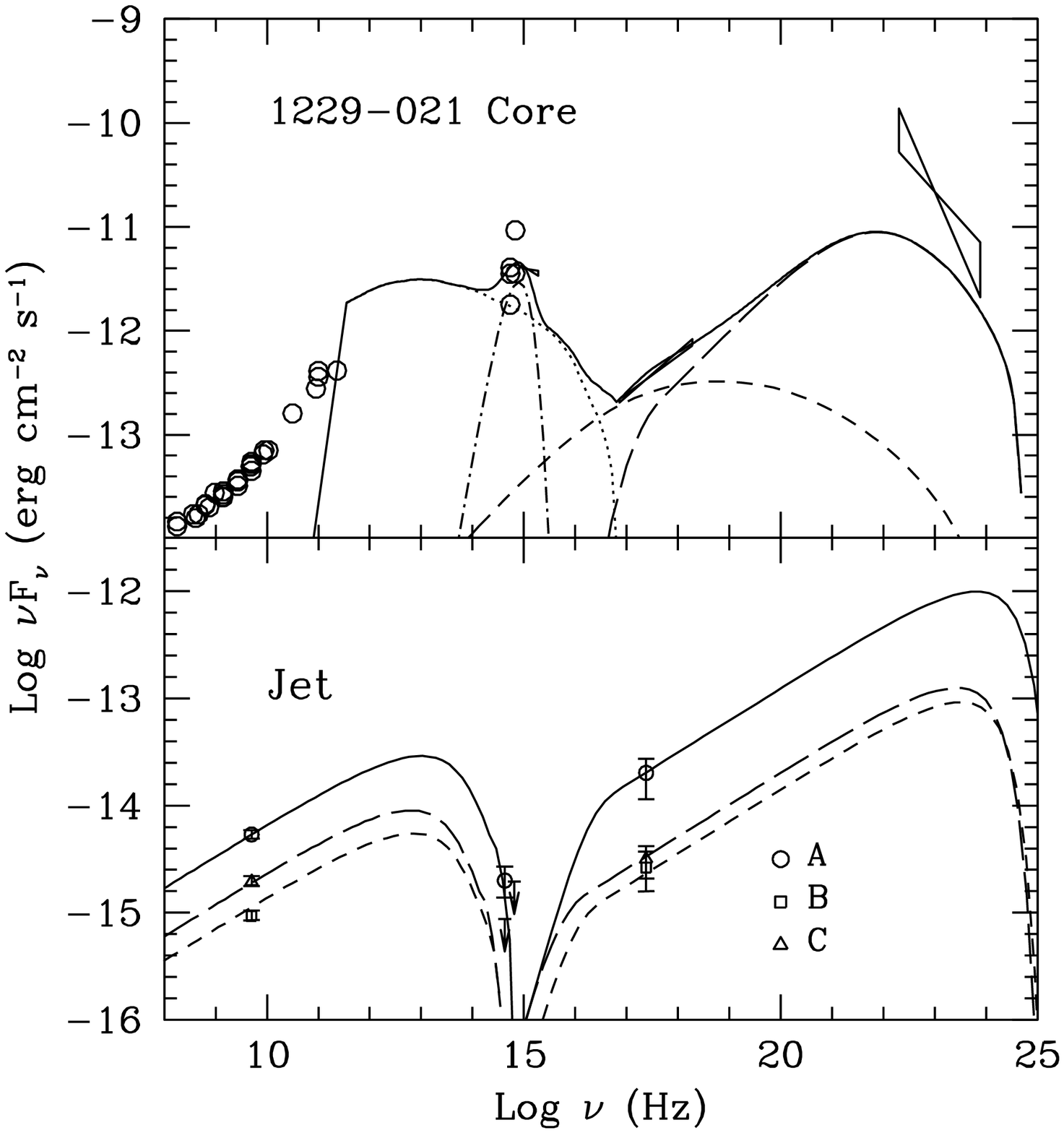}}
\caption{}
\end{figure}

\begin{figure}[]
\figurenum{2d}
\noindent{\plotone{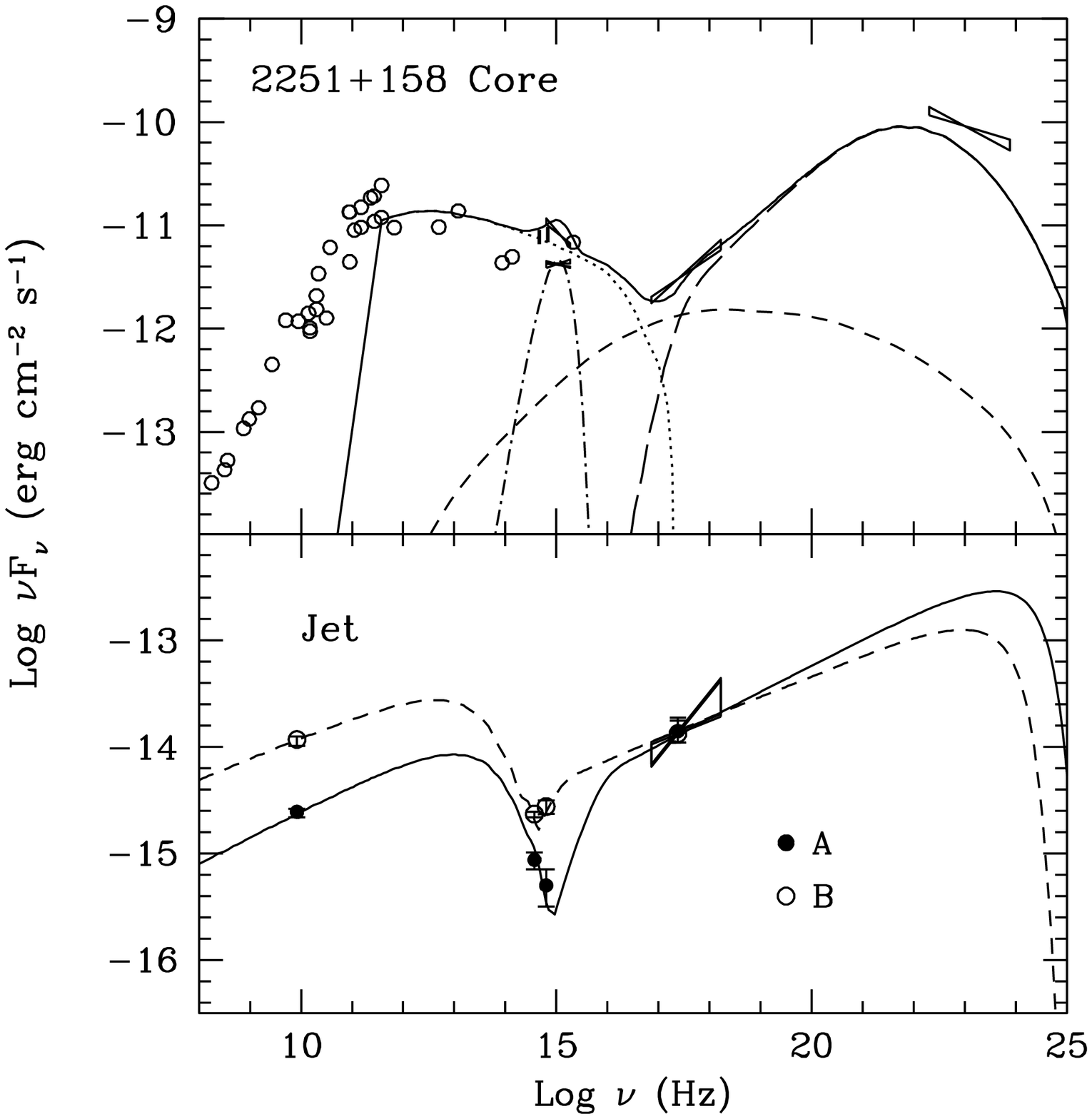}}
\caption{}
\end{figure}

\begin{figure}[]
\figurenum{3}
\noindent{\plotone{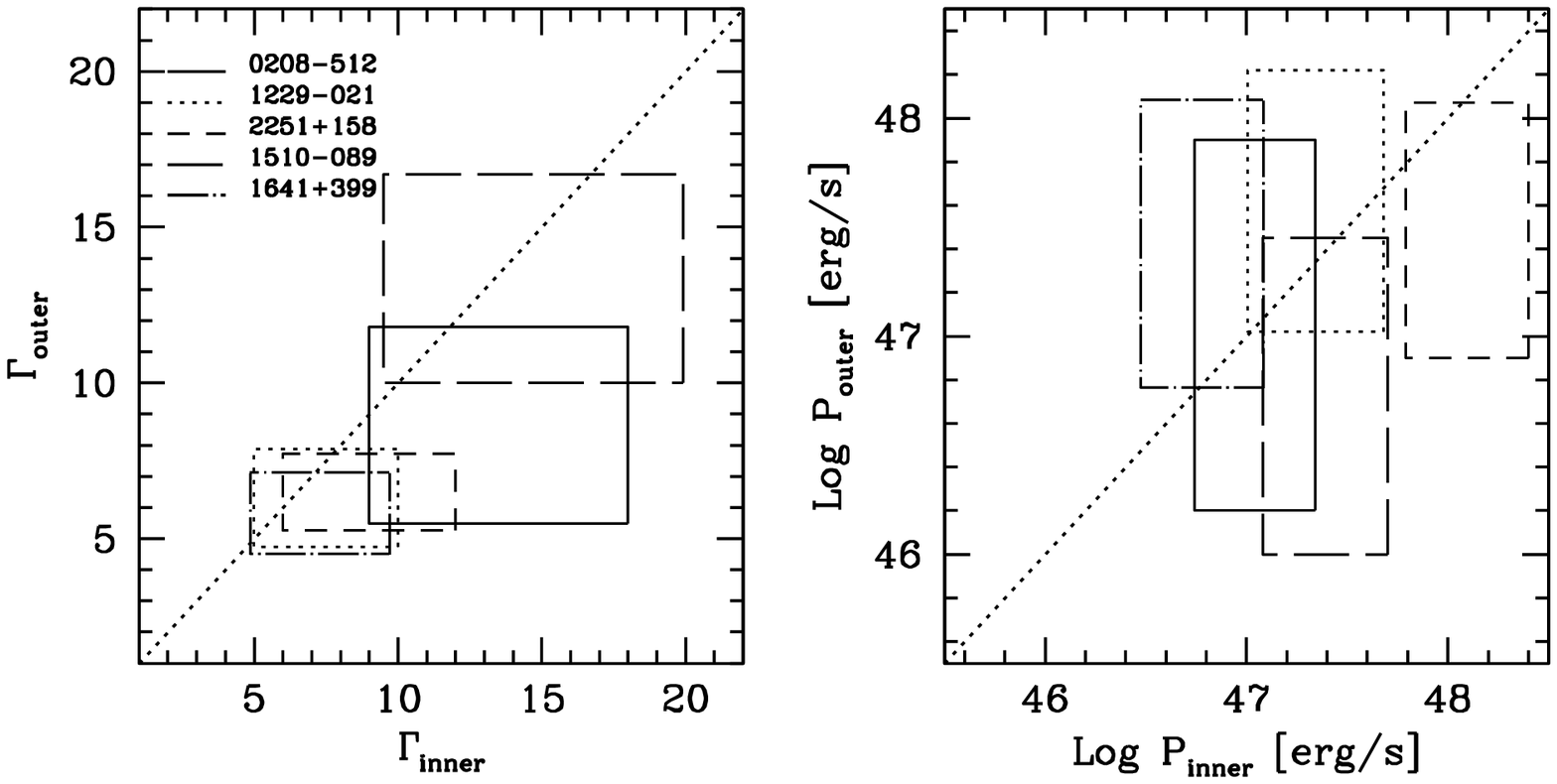}}
\caption{}
\end{figure}

\end{document}